\documentclass[journal]{IEEEtran}
\usepackage{amsmath,amsfonts}
\usepackage{algorithmic}
\usepackage{algorithm}
\usepackage{array}
\usepackage{textcomp}
\usepackage{stfloats}
\usepackage{url}
\usepackage{verbatim}
\usepackage{graphicx}
\usepackage{cite}
\usepackage[T1]{fontenc}
\usepackage{booktabs} 
\usepackage{subfig}

\usepackage{caption}
\usepackage{wrapfig}
\usepackage{enumerate}
\usepackage{gensymb}
\usepackage{color}
\usepackage{amsmath}
\usepackage{dcolumn}
\usepackage{multirow}
\usepackage{amssymb}
\usepackage{flushend}

\begin{document}

\title{Exploring quantum sensing for fine-grained liquid recognition}

\author{IEEE Publication Technology,~\IEEEmembership{Staff,~IEEE,}
\author
{Yuechun Jiao, 
Jinlian Hu, 
Zitong Lan, 
Fusang Zhang, 
Jie Xiong, 
Jingxu Bai, 
Zhaoxin Chang, 
Yuqi Su,\\ 
Beihong Jin, 
Daqing Zhang, 
Jianming Zhao, 
Suotang Jia\\
}

\thanks{This paper was produced by the IEEE Publication Technology Group. They are in Piscataway, NJ.}
\thanks{Manuscript received April 19, 2021; revised August 16, 2021.}}

\maketitle

\begin{abstract}
Recent years have witnessed the use of pervasive wireless signals (e.g., Wi-Fi, RFID, and mmWave) for sensing purposes. Due to its non-intrusive characteristic, wireless sensing plays an important role in various intelligent sensing applications. However, limited by the inherent thermal noise of RF transceivers, the sensing granularity of existing wireless sensing systems are still coarse, limiting their adoption for fine-grained sensing applications. In this paper, we introduce the quantum receiver, which does not contain traditional electronic components such as mixers, amplifiers, and analog-to-digital converters (ADCs) to wireless sensing systems, significantly reducing the source of thermal noise. By taking non-intrusive liquid recognition as an application example, we show the superior performance of quantum wireless sensing. By leveraging the unique property of quantum receiver, we propose a novel double-ratio method to address several well-known challenges in liquid recognition, eliminating the effect of liquid volume, device-target distance and container. 
We implement the quantum sensing prototype and evaluate the liquid recognition performance comprehensively. The results show that our system is able to recognize 17 commonly seen liquids, including very similar ones~(e.g., Pepsi and Coke) at an accuracy higher than 99.9\%. For milk expiration monitoring, our system is able to achieve an accuracy of 99.0\% for pH value measurements at a granularity of 0.1, which is much finer than that required for expiration monitoring.
\end{abstract}


\section{Introduction}
\IEEEPARstart{W}{ireless} technologies have changed our world in every aspect in the last two decades. Wi-Fi and cellular connections have become an indispensable part of our everyday life. There are more than 400 million wireless devices in the US alone~\cite{url4} and 92.3\% of Internet users worldwide access the Internet using wireless connections~\cite{url3}. We witnessed a huge success in utilizing wireless technologies for data communication. In recent years, a new research area called wireless sensing has emerged. Instead of data communication, researchers utilize wireless signals in our surrounding environment for sensing purposes. For example, Wi-Fi signals are utilized to sense human contexts such as hand gestures~\cite{whole2013} and vital signs~\cite{From2018}. The basic principle behind wireless sensing is that human movements induce signal variations and by analyzing the signal variations, target movement information can be obtained. Besides sensing human contexts, wireless sensing has also been leveraged to sense non-human objects~(e.g., material sensing) and even the environment~(e.g., temperature and humidity sensing). The unique contact-free nature~(i.e., does not need to be in contact with the target) makes wireless sensing appealing in a large range of real-world applications. 

Although promising, there are still several limitations associated with wireless sensing. One critical limitation is the coarse granularity. 
Take Wi-Fi sensing as an example, when motion displacement is less than 0.3~cm, the motion induced signal variation is too small and submerged in noise without being detected. In liquid sensing, Wi-Fi signal is capable of differentiating between water and orange juice. However, it is challenging for Wi-Fi signals to achieve fine-grained differentiation between two orange juice of different brands. 
This is because the sensing granularity is heavily affected by the inherent hardware thermal noise. For Radio Frequency~(RF) devices, electric components such as mixers, amplifiers, and Analogue-to-Digital-Converters~(ADCs) unavoidably introduce thermal noise to the signal and the introduced noise becomes the fundamental factor limiting the sensing granularity.

With this fundamental limit in mind, we propose to involve the quantum receiver in the ecosystem of wireless sensing. The Rydberg atom-based quantum receiver~\cite{ Fan2015} is composed of atoms with electrons excited to high energy levels~\cite{gallagher1994rydberg}. Different from conventional RF receivers which incur thermal noise, the quantum noise is much smaller~\cite{Wang:23}.  
The significantly smaller noise can greatly boost the sensing granularity of wireless signals. Also, the quantum receiver is capable of receiving wireless signals in a much larger frequency range~(e.g., 10~MHz to 1~THz)~\cite{Jiao2017, Jing2020}. In contrast, for traditional RF receivers, one device works in a much smaller frequency range and even antennas need to be redesigned for efficient reception of signals in different frequency bands.

To demonstrate the superior sensing performance, we employ liquid sensing as the application example. Liquid sensing plays an important role in a lot of disinclines such as quality inspection and expiration detection. Take milk expiration as an example. Expired milk causes a lot of digestive issues including bloating, diarrhea, and stomach pain. On the other hand, 16\% fresh milk is wasted because people are unsure whether it has expired or not. Conventional methods for milk expiration detection usually rely on expensive (several thousands of dollars) specialized equipment, such as spectrometers~\cite{Mass2012}. These methods further require collecting liquid samples and placing them in specialized containers followed by a complicated calibration procedure~\cite{LiquID2018}. We observe two unique advantages of applying a quantum receiver for liquid sensing. 

\begin{figure*}[t]
    \centering
\includegraphics[width=0.9\linewidth,height=6.5cm]{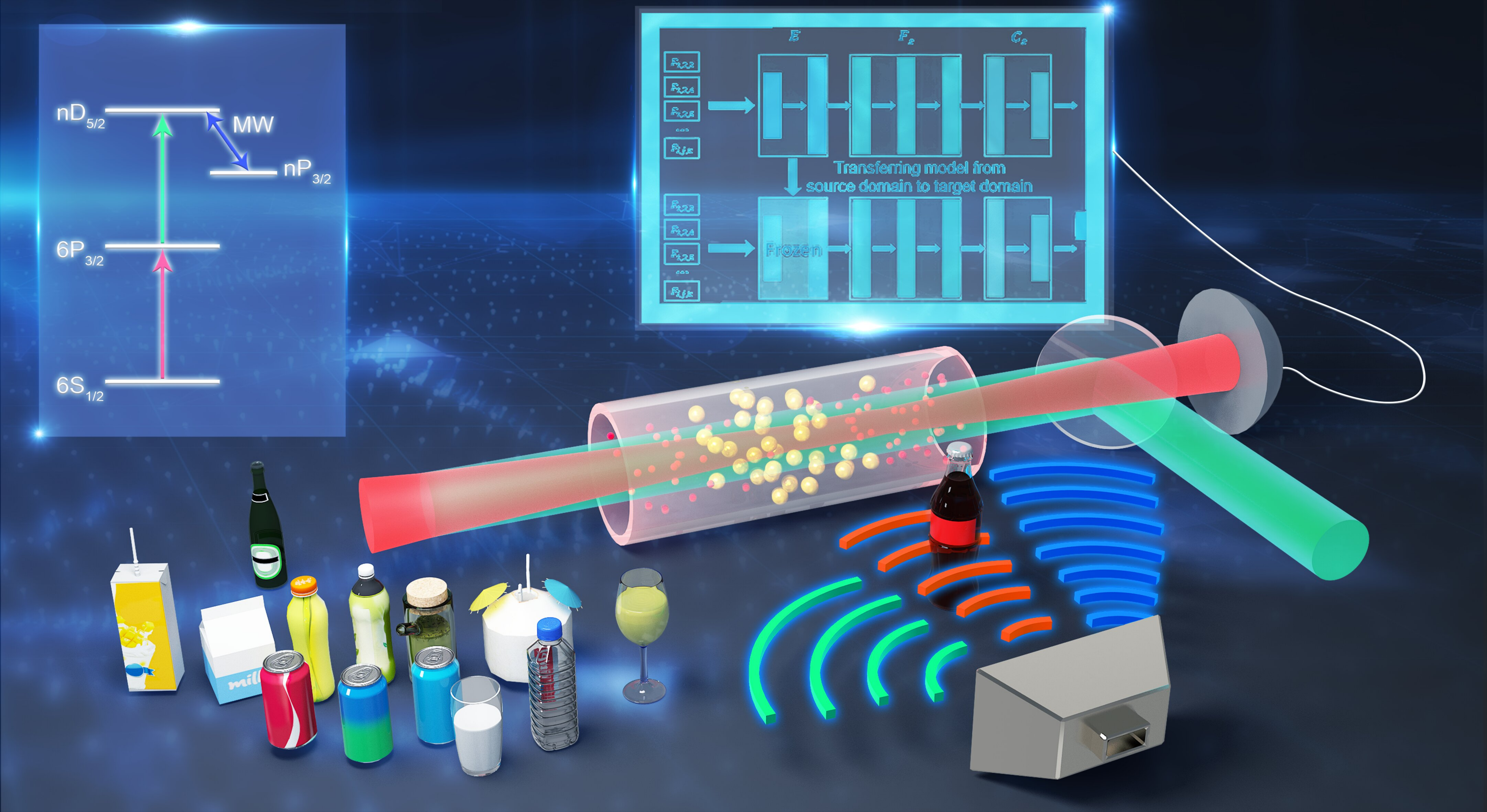} %
    \caption{Illustration of quantum-based liquid sensing system.}
    \label{fig:setup}
\end{figure*}

\begin{enumerate}
    \item 
    \textbf{Non-intrusive:} Compared to traditional sensing methods, quantum wireless sensing does not require taking any liquid samples out of the original case. 	
    \item 
    \textbf{Finer sensing granularity:} Compared to conventional wireless sensing, the sensing granularity is much finer and accordingly the accuracy of quantum wireless sensing is much higher. 
\end{enumerate}

While promising, there are two well-known challenges associated with wireless liquid sensing. 
The first challenge is liquid volume dependency and container dependency. This is because the rationale behind wireless liquid sensing is that wireless signals penetrate through the liquid and the signals experience different amounts of variations inside different liquids. However, this is only valid when the liquid volume is the same. The signal variations caused by a small bottle of water and a large bottle of water are obviously different. Another factor affecting signal variation is the container. Different containers also cause different amounts of signal variations, making liquid recognition challenging. Removing the volume and container dependency is non-trivial.

The second challenge is signal frequency dependency. We observe that the liquid sensing performance is sensitive to signal frequency. For example, while the difference of one liquid feature we use for liquid recognition at 5.04~GHz between Pepsi and Coke is merely 1.2$\%$, the difference at 16.31~GHz is 68.8$\%$. What makes it more challenging is that different liquids are sensitive to different frequencies~\cite{deng2019noninvasive}. Thus, the sensing performance is highly dependent on signal frequency and one frequency does not work well for all liquids. 
	
In this paper, we leverage the unique property of the quantum receiver, as shown in Figure~\ref{fig:setup}, i.e., one quantum receiver is capable of receiving signals in a large frequency range to address the above two challenges.  We model the theoretical relationship between the liquid and the signal strength. The signal strength is affected by various factors including signal propagation distance, transmission power, liquid material type, liquid volume, and container. To remove the effect of other factors and only keep the effect of liquid material, we propose a double-ratio method. We first employ the signal ratio of two different frequencies to remove the effect of signal propagation distance and transmission power. We further exploit the signal of a third frequency to perform a logarithmic ratio operation to remove the influence of container. After removing the effect of all interfering factors, we achieve contact-free liquid sensing without a need of taking liquid samples out from their original case/container for the first time. 
	
To tackle frequency dependency, we fully leverage the capability of the quantum receiver to receive signals at multiple different frequencies. The frequency diversity effectively gets rid of the effect of frequency dependency and enhances distinguishability. With rich information from multiple frequencies, we design a very lightweight machine learning model to
achieve fine-grained liquid material sensing. To adapt the proposed machine learning model for different applications~(e.g., liquid recognition and milk pH level measurement), we keep the feature extraction layers of the machine learning model untouched and only refine the classifier layer to meet the specific objective of each application. 
	
Our experiments show that the proposed system is able to accurately distinguish 17 commonly-seen liquids at an accuracy higher than 99.9\%. The 17 liquids include very similar ones such as (i) green and black tea, (ii) orange juice of different brands and (iii) Pepsi and Coke, demonstrating an extremely fine-grained sensing granularity. 
In the liquid concentration experiment, the proposed quantum sensing can detect a 1\% sugar/salt concentration change. For milk expiration monitoring, the proposed system can achieve an accuracy of 99.0\% in detecting a pH level change of 0.1, fine enough to detect milk expiration.

\section{Quantum primer}
\subsection{Rydberg atom}
\begin{figure}[!htbp]
    \centering
    \includegraphics[width = 0.8\linewidth]{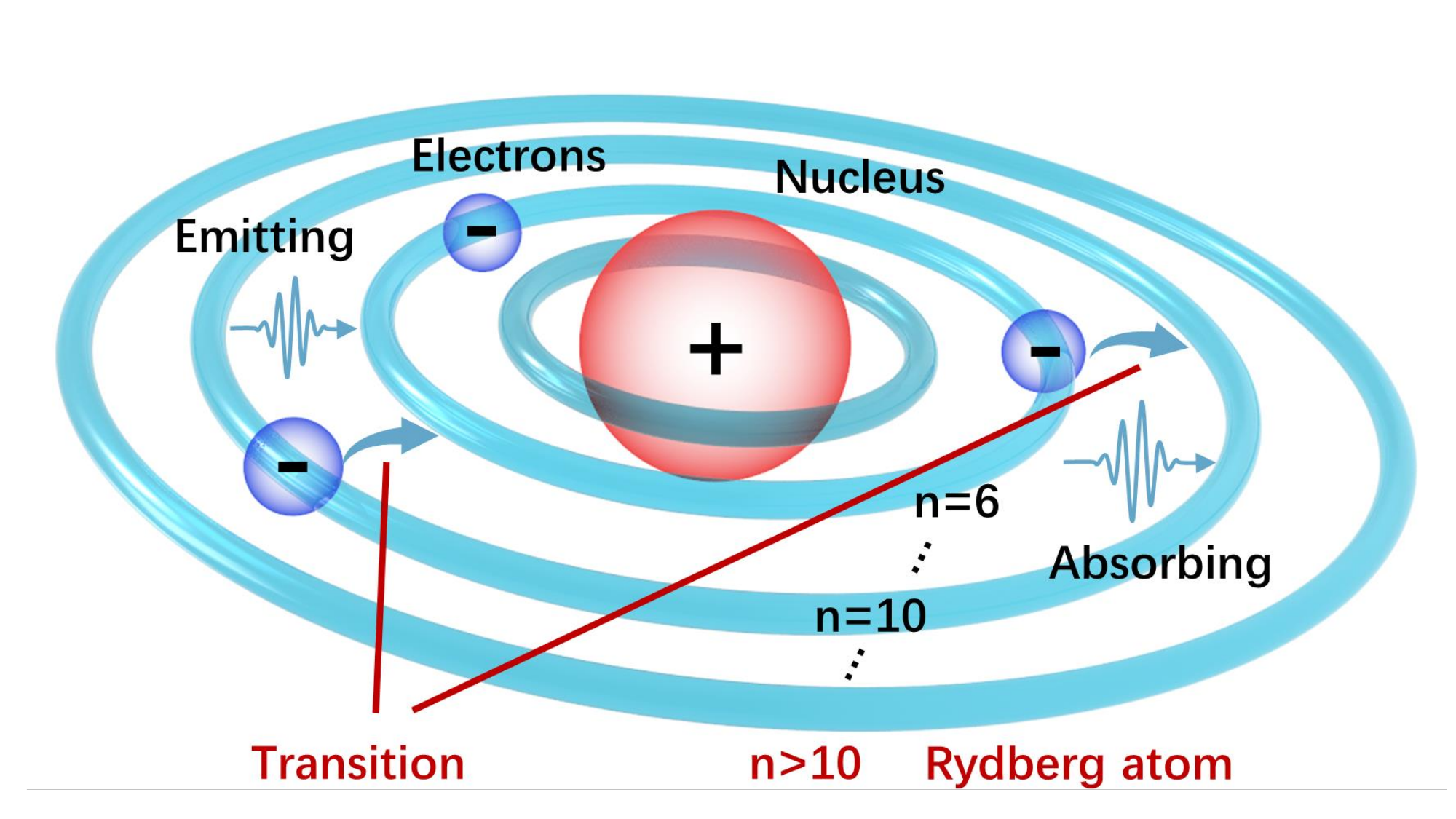}
    \caption{Atom electron transitions and Rydberg atom.}
    \label{fig:Rydberg atom}
\end{figure}

The smallest unit of matter is an atom, which contains one nucleus and multiple electrons as shown in Figure~\ref{fig:Rydberg atom}. The electromagnetic force of the nucleus binds the electrons to a certain set of orbits. The electrons on different orbits have different energy levels. 
An electron can release energy to travel to lower energy levels or absorb energy to travel to higher energy levels, as shown in Figure~\ref{fig:Rydberg atom}. Electrons have the ability to absorb or release energy through the reception or emission of electromagnetic radiation. An atom with one or more electrons at a high energy level (i.e., energy level index~(n) $>$ 10)  is known as a \textbf{Rydberg atom}.

\subsection{Quantum receiver based on Rydberg atoms}

To receive radio frequency~(RF) such as Wi-Fi signals, a quantum receiver based on Rydberg atoms can be used. The main component of a quantum receiver is a vapor cell filled with atoms. 
We then employ laser beams to trigger electrons of the atoms to high energy levels, i.e., these atoms become Rydberg atoms. To use a quantum receiver to receive RF signals of a particular frequency, we need to trigger the electrons to a specific energy level. 
For example, the transition from energy level $L_1$=67$P_{3/2}$ to energy level $L_2$= 66$D_{5/2}$ induces RF signals of 2.4~GHz frequency which is the Wi-Fi frequency.

To measure the RF signal strength, we observe that the reception of RF signal results in a split of the spectrum of the trigger laser beam and the signal strength is linearly related to the amount of split. The spectrum of the laser beam can be obtained using a photodetector. The quantitative relationship between the peak difference of the spectrum and the strength of the RF signal can be derived using the well-known electromagnetically induced transparency and Autler–Townes (EIT-AT) effect~\cite{sedlacek2012microwave}. 

\section{Quantum-based liquid sensing}

Now we utilize a quantum receiver based on Rydberg atoms to sense liquid type. A diagram of our sensing system is shown in Figure~\ref{fig:setup}. Two lasers of wavelength 510~nm and 852~nm are used to excite Cs atoms to the Rydberg state for RF signal reception.  
A photodetector is used to measure the laser spectrum and obtain the strength of the RF signal. The detected signal information is stored in a data acquisition board~(NI USB-6363). After the data acquisition board receives the raw signal samples, it forwards them to a laptop through a Wi-Fi connection for processing. When signals of different frequencies pass through a liquid, we obtain the signal amplitude change measured at the quantum receiver.
The signals are first processed to remove the effect of container size. Then we fed the processed signals into a multi-task transfer learning model, which consists of two hidden layers to capture the common features and three task-specific layers to conduct feature processing. 
The quantum wireless sensing system is capable of recognizing liquid types, as well as estimating liquid concentration and pH level.

\subsection{The relationship between liquid permittivity and signal strength}
Note that the liquid is in its original container and placed between the RF transmitter and the quantum receiver. We do not need to pour the liquid into a specific container for recognition as conventional methods do. 
Electromagnetic waves are refracted and reflected when they reach the boundary between two media (e.g., the boundary between air and container). 
The signal is attenuated when it propagates through the liquid. 
The amount of attenuation is determined by the attenuation factor, $\beta_f$. 
The attenuation factor is defined as the amount of propagation distance inside the liquid to decay the electromagnetic field's strength to $\frac{1}{e}$ of its initial level~\cite{LiquID2018,feynman2011feynman}. For convenience of representation, $\beta_f$ is denoted as $\beta_f $= $\frac{2\pi}{\lambda} \sqrt{\frac{\mathbb{R}_f}{2}(\sqrt{1+\frac{\mathbb{I}_f}{\mathbb{R}_f}}-1)}$ and $\lambda$ represents the RF signal's wavelength. 
The real and imaginary components of the liquid's complex permittivity are denoted by $\mathbb{R}_f$ and $\mathbb{I}_f$, respectively. Note that the permittivity also varies with signal frequency.  
Given a specific signal frequency, the liquid's complex permittivity, which can be utilized as a feature to identify the liquid, is the only factor on which the attenuation factor depends. To identify the liquid type, assume that we send a signal with strength $S$ from transmitter, we measure the strength of the received signal at the quantum receiver, which can be expressed as $E_f$ = A $\cdot \kappa \cdot e^{-\beta_f d} \cdot P \cdot S$, where $A$ is the attenuation of the signal in the air, $d$ is the propagation distance of the signal in the liquid, $P$ is the gain of the quantum receiver and $\kappa$ is the attenuation induced by the container. 

\subsection {Signal ratio of multiple frequencies}
Based on the above derivation, we can see that the energy of the received signal is influenced by many factors related to the transmission signal and the environment. To remove these influences, we leverage the unique advantage of a quantum receiver, i.e., it can receive signals of dramatically different frequencies with a single receiver. 
Without loss of generality, for the same liquid, we first send two signals of frequencies $f_i$ and $f_j$, and the quantum receiver only needs to excite the atoms to corresponding energy levels to receive signals at these two frequencies. In this way, we can capture the signal penetrating through the liquid at the quantum receiver, which are given by:
\begin{equation}\label{eq:3}
	E_{f_{i}} = A \cdot \kappa \cdot e^{-\beta_{f_{i}} \cdot d} \cdot P \cdot S.
\end{equation}
By taking the ratio of the two signals, other parameters are removed and only the attenuation factor is preserved: 
\begin{equation}\label{eq:5}
    \frac{E_{f_i}}{E_{f_j}} = \frac{A \cdot \kappa \cdot e^{-\beta_{f_i} d} \cdot P \cdot S}{A \cdot \kappa \cdot e^{-\beta_{f_j} d} \cdot P \cdot S} = e^{-d(\beta_{f_i}-\beta_{f_j})}.
\end{equation}

We observe that Equation~\ref{eq:5} yields a representation only related to the liquid width~($d$) and the attenuation factor~($\beta_{f_i}$ and $\beta_{f_j}$). Note that the attenuation factor is related to the material property, i.e., the dielectric constant. 
In real-world settings, we still need to remove the effect of container to make our method container-independent. To achieve this, we further utilize signals of more frequencies. 
We propose a novel double-division method to address the issue of container diversity. 
Assuming that we have a third signal of frequency $f_k$ penetrating through the liquid, we can now obtain two pairs of signal expressions containing the liquid width, for which we take the $ln(.)$ operation as 
$ln(\frac{E_{f_{i,j}}}{E_{f_k}})$= $-d(\beta_{f_{i,j}}-\beta_{f_k})$.
Through another division operation,  
we can remove the container width~(size) and obtain the representation only related to the dielectric constant of the liquid, denoted as $F_{i,j,k}$:
\begin{equation}\label{eq:8}
    \begin{aligned}
        F_{i,j,k} & = ln(\frac{E_{f_i}}{E_{f_j}}) / ln(\frac{E_{f_j}}{E_{f_k}}) = -d(\beta_{f_i}-\beta_{f_j}) / -d(\beta_{f_j}-\beta_{f_k}) \\
                  & = (\beta_{f_i}-\beta_{f_j})/(\beta_{f_j}-\beta_{f_k}).
    \end{aligned}
\end{equation}

Essentially, the Rydberg-based quantum receiver can select a large number of frequencies for sensing. 
Under each combination of three frequencies, one feature vector can be extracted. With multiple combinations, more accurate and robust liquid recognition can be achieved.

\begin{figure*}[!t]
    \centering
    \includegraphics[width = 0.95\linewidth,height=6cm]{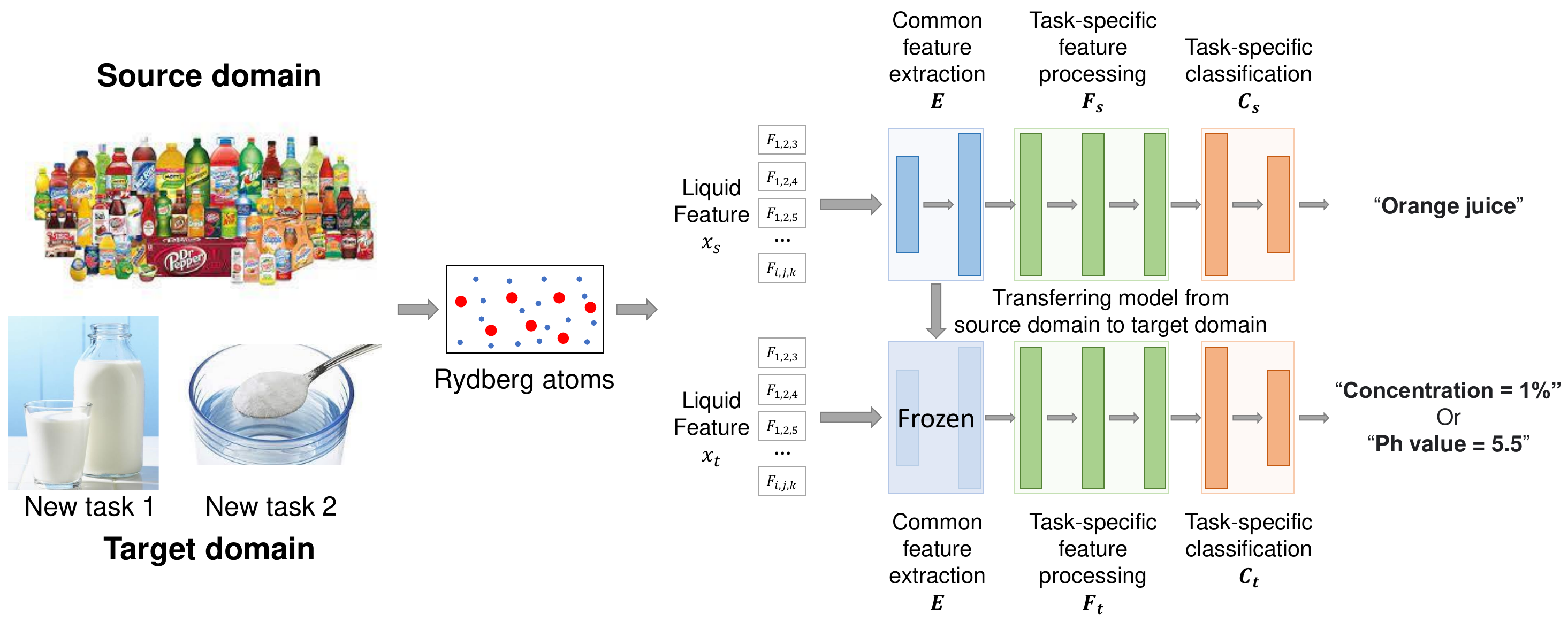}
    \caption{Quantum feature based multi-task transfer learning framework for liquid sensing.}
    \label{fig:system_ar}
\end{figure*}

\subsection{Multi-task transfer learning framework for liquid sensing}

Based on the proposed multi-frequency feature, we propose a transfer learning-based framework to perform liquid sensing. The main task is to identify the liquid type. Meanwhile, our model can be easily extended to deal with new sensing tasks with a small amount of extra data and training. For example, our system can be extended to measure the concentration of sugar in water and the pH level of milk for milk expiration monitoring.

The multi-task transfer learning framework is illustrated in Figure~\ref{fig:system_ar}. It consists of a source domain and a target domain. The source domain trains the quantum feature data collected from different liquids~(e.g., Coke, juice, bear, milk, etc) and the target domain uses a smaller number of new target data (e.g., expired milk with different pH levels) to train a classifier that can estimate the pH level of the milk. 
Basically, these two domains share similar structure, which consists of an input layer, multiple hidden layers and an output layer. These layers can be further divided into common feature extraction layers $E$, task-specific feature processing layers $F$ and task-specific classification layer $C$. Common feature extraction layers extract liquids' common feature from signal data collected at the quantum receiver. Task-specific feature processing layers process these features according to the task. The task-specific classification layer changes its output dimensions to match the number of classes in a specific task. 

To extend the classifier to a new target domain, the common feature extraction layers can be directly reused. To be specific, their weights are fixed in the training process to help reduce the number of parameters that need to be updated for the new task. The task-specific feature processing layers and their parameters are updated in the training process for the new task. The dimension of the task-specific classification layer is set to be the number of classes in the new task. Mathematically, we compute $y_s = E(x_s)\cdot F_s \cdot C_s$, where $x_s$ denotes the input liquid feature vector from the source domain. We then minimize the loss function $\mathcal{L}(y_s,\tilde{y}_s)$ between the model prediction $y_s$ and the ground truth label $\Tilde{y}_s$ and optimize the parameters of all three parts including $E(.)$, $F_s$ and $C_s$ in the source domain. In the target domain, we compute $y_t =  E(x_t)\cdot F_t \cdot C_t$ and compare with the ground truth to optimize the parameters of $F_t$ and $C_t$.

\section{Experiment Results}

\subsection{Experimental setup}
\begin{figure*}[t]
    \centering
    \includegraphics[width=1\linewidth]{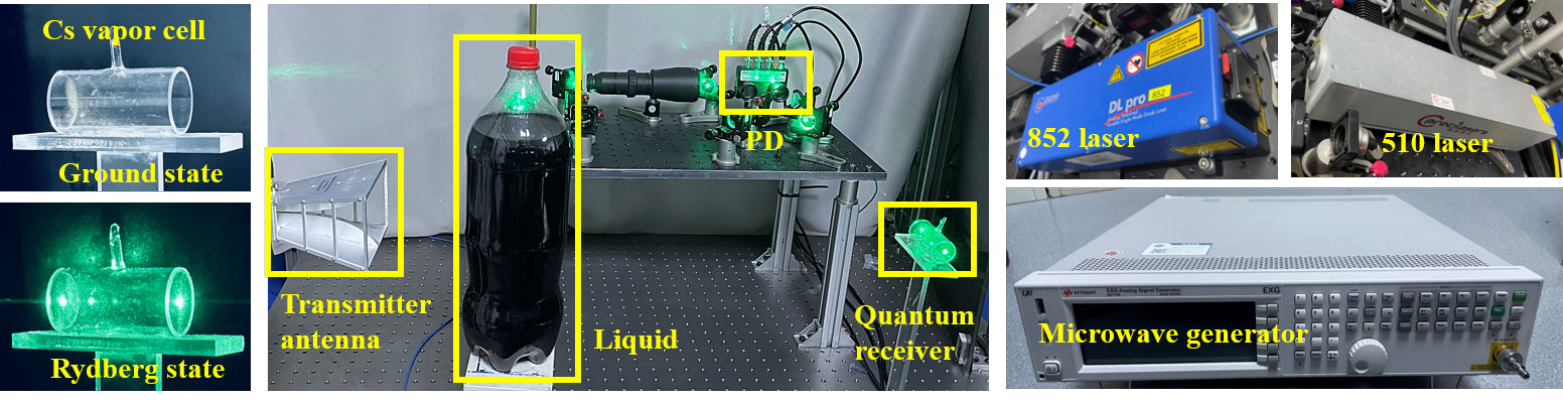}
    \caption{Experiment setup in a real environment.}
    \label{fig:real_apparatus}
\end{figure*}

The experiment setup in a real environment is shown in the middle of Figure~\ref{fig:real_apparatus}. 
The key component of the proposed quantum wireless sensing system is an atomic vapor cell as shown in Figure~\ref{fig:real_apparatus} (left). In our experiment, the size of the cell is $2.5~cm \times 5.0~cm$, which is filled with Cs atoms. The lower left of Figure~\ref{fig:real_apparatus} shows a vapor cell with Cs atoms excited to Rydberg atoms. The Rydberg atom-based quantum system typically has a ground state (e.g., $|6S_{1/2}, F = 4\rangle$), an intermediate state (e.g., $|6P_{3/2}, F^\prime = 5\rangle$) and a Rydberg state (e.g., $|r\rangle = nD_{5/2}$). The atomic level is normally denoted as $nl_{J}$, where $n$ is principal quantum number, $l$ is orbital angular momentum and $J$ is the total electron angular momentum. A probe laser ($\lambda_p$ = 852~nm) excites the transition of $|6S_{1/2}, F = 4\rangle\to |6P_{3/2}, F^\prime = 5\rangle$ with a power of  150 $\mu$W, and the second laser ($\lambda_c$ = 510~nm), counter-propagating with the probe laser, is used to achieve the Rydberg transition to the state of $|r\rangle = nD_{5/2}$ from $|6P_{3/2}, F^\prime = 5\rangle$ with a power of 12 mW. The second coupling laser also has the function of  establishing the electromagnetically induced transparency (EIT) for RF signal measurement~\cite{Mohapatra2007}. 
The EIT signal is obtained by measuring the spectrum of the probe laser using a photodiode (PD). 
The wireless signals of ten different frequencies (i.e., 5.04~GHz, 5.71~GHz, 6.49~GHz, 7.43~GHz, 8.56~GHz, 9.93~GHz, 11.61~GHz, 13.69~GHz, 14.92~GHz and 16.31~GHz) are generated by a signal generator (KEYSIGHT N5183B) and emitted using a broadband antenna (A-info LB-20180-SF). The transmitter antenna is placed 70~cm away from the atom cell, and the target liquid is placed between the transmitter antenna and the receiver.


\subsection {Transfer learning model implementation}
In the source domain, we use Adam optimizer and exponential learning rate decay as our optimization techniques. The model is trained for 100 epochs with a learning rate = $10^{-3}$. For the last layer, it outputs the prediction result, which is one of the seventeen liquids in the source domain.

After training in the source domain, we transfer the model to train it in the target domain. We consider the first two hidden layers as the common feature extraction layers $E$ and their parameters are fixed during the training process. The following three hidden layers are task-specific feature processing layers $F$ and their parameters are updated during the training process. The dimension of the output layer is 15 and 17 for salt (sugar) concentration estimation and milk pH value sensing, respectively. We set training epochs = 50 and learning rate = $10^{-4}$ in these two tasks. Note that we use the parameters of a well-trained model in the source domain, so the model can converge to the optimal state faster when trained in the target domain, and the training epoch required is less.

\subsection{Benchmark experiment on liquid recognition}

\begin{figure*}[!htbp]
    \centering
\subfloat[Confusion matrix of liquid identification accuracy] {
\label{fig:confusion}
\includegraphics[width=0.55\linewidth,height=4.3cm]{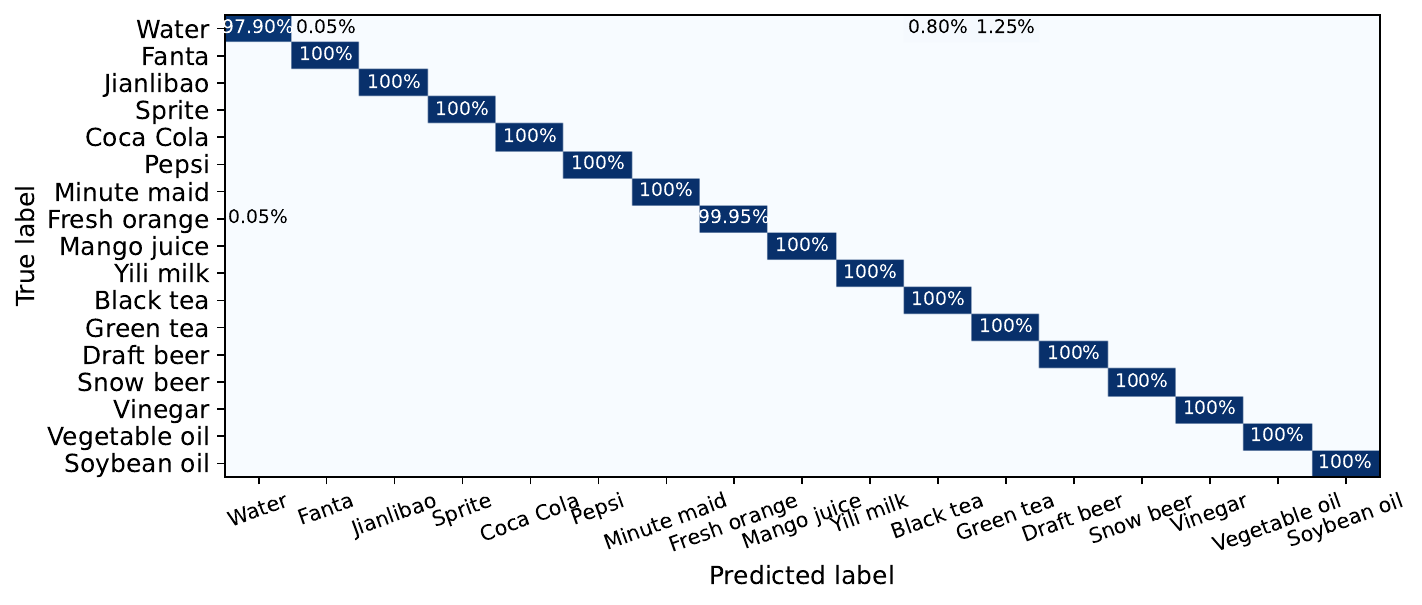}
}
\subfloat [Comparison of identification accuracy with Wi-Fi and USRP] {
\label{fig:compare}
\includegraphics[width=0.45\linewidth,height=4.3cm]{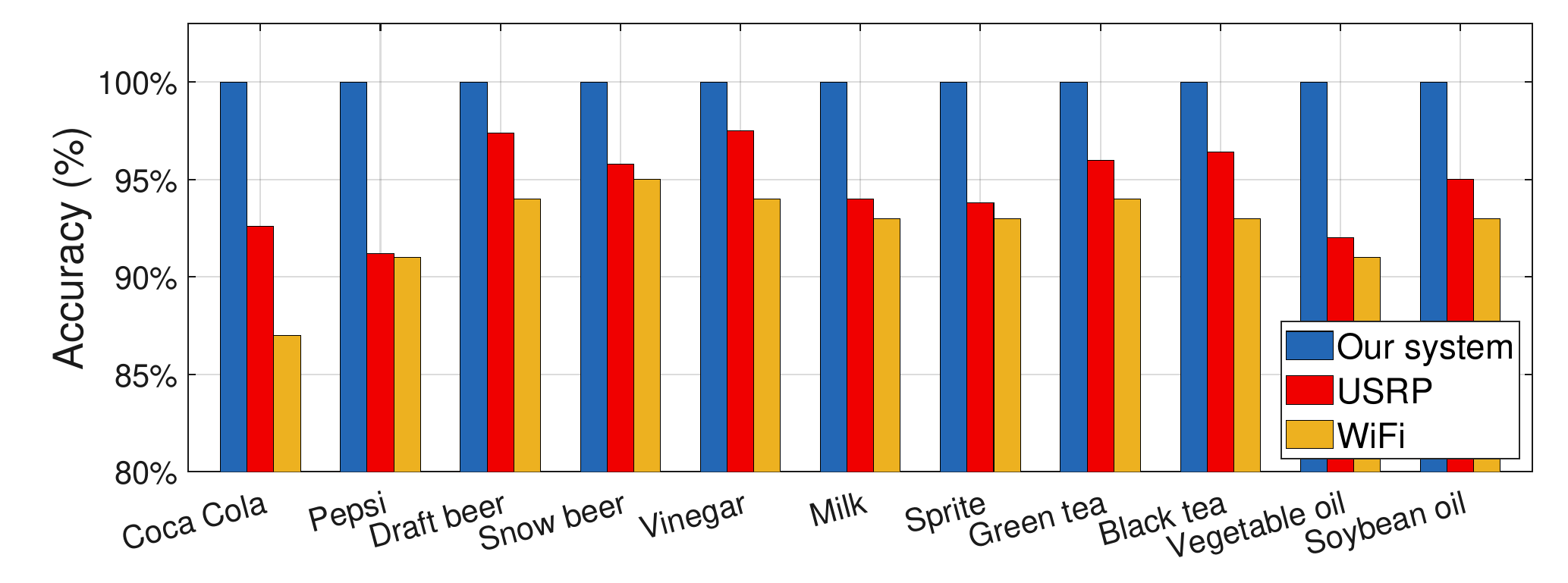}
}
\caption{Performance of differentiating 17 commonly seen liquids.}
\end{figure*}

To verify the effectiveness of the proposed quantum sensing system, we employ seventeen commonly-seen liquids (i.e., water, Fanta,  Sprite, Coca Cola, Pepsi, tea, orange juice, mango juice, milk, etc.) and differentiate these liquids without opening the cases/bottles. Note that for Pepsi and Coke, we employ the plastic bottle versions as RF signals can hardly penetrate metal cases. All the liquids are bought from a supermarket with their original cases/bottles. We perform data collection for 17 liquids 3 times a day for 5 consecutive days. 
The confusion matrix in Figure~\ref{fig:confusion} shows the results of liquid recognition. We can see that the average accuracy reaches 99.9\%. For similar liquids such as Pepsi and Coke, the proposed sensing system is able to differentiate them at an accuracy of 100\%.


\subsection{Quantum wireless sensing vs. conventional sensing}

To compare the performance of the proposed quantum wireless sensing with conventional wireless sensing, we employ the widely used commodity Intel 5300 Wi-Fi card and dedicated low-noise Universal Software Radio Peripheral (USRP) B210 as the receiver for conventional wireless sensing. For a fair comparison, we use the same container in all experiments. We also utilize the same recognition model for training. Figure~\ref{fig:compare} shows the recognition accuracy of the proposed system, commodity Wi-Fi sensing system and low-noise USRP-based Wi-Fi sensing. We choose similar liquids~(e.g., draft beer and snow beer, green tea and black tea, etc.) to show the recognition results. We can see that for similar liquids, the proposed system, conventional Wi-Fi system and low-noise USRP achieve an accuracy of 100\%, 92.5\% and 94.7\%, respectively, demonstrating the superior performance of quantum wireless sensing.

\subsection{Performance of fine-grained solution concentration sensing} 
Quantum wireless sensing can sense fine-grained changes in solution concentration. In this subsection, we evaluate the accuracy of liquid concentration sensing. Similar to the liquid recognition experiments, we measure the signal amplitude at the quantum receiver and use measurements at multiple frequencies to recognize solution concentration. 
We dissolve salt~(sugar) in water, and vary the concentration from 5\% to 15\% at a step size of 1\%.   
The results in Figure~\ref{fig:salt}
show that for a concentration change equal or larger than 2\%, our system can 100\% detect the change. For a 1\% concentration change, our system can still achieve a high detection rate of 94.8\%.  

\subsection{Performance of milk pH level measurement}

\begin{figure*}[!htbp]
\centering
\subfloat[Concentration estimation] {
\label{fig:salt}
\includegraphics[width=0.3\linewidth]{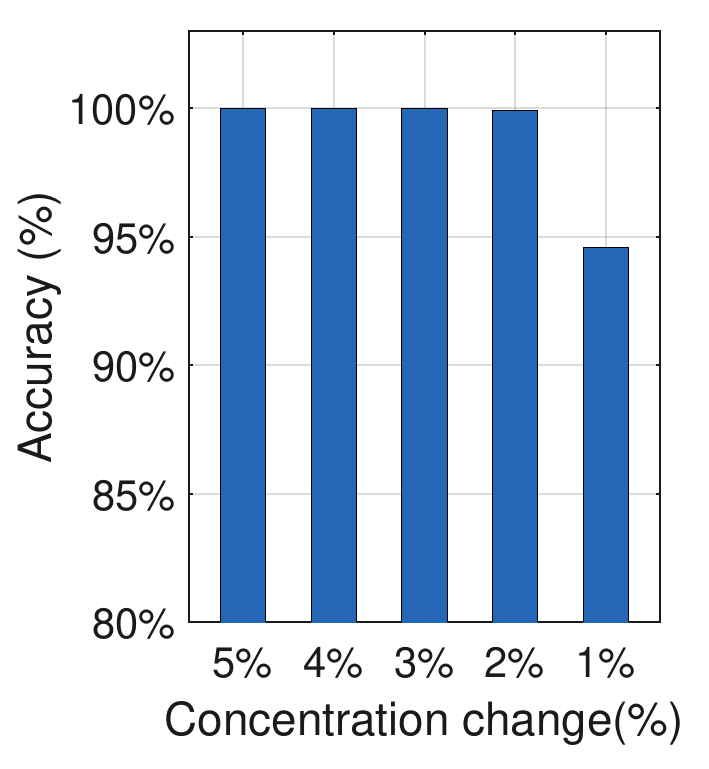}
} 
\quad
\subfloat[pH level measurement] {
\label{fig:ph_level}
\includegraphics[width=0.55\linewidth]{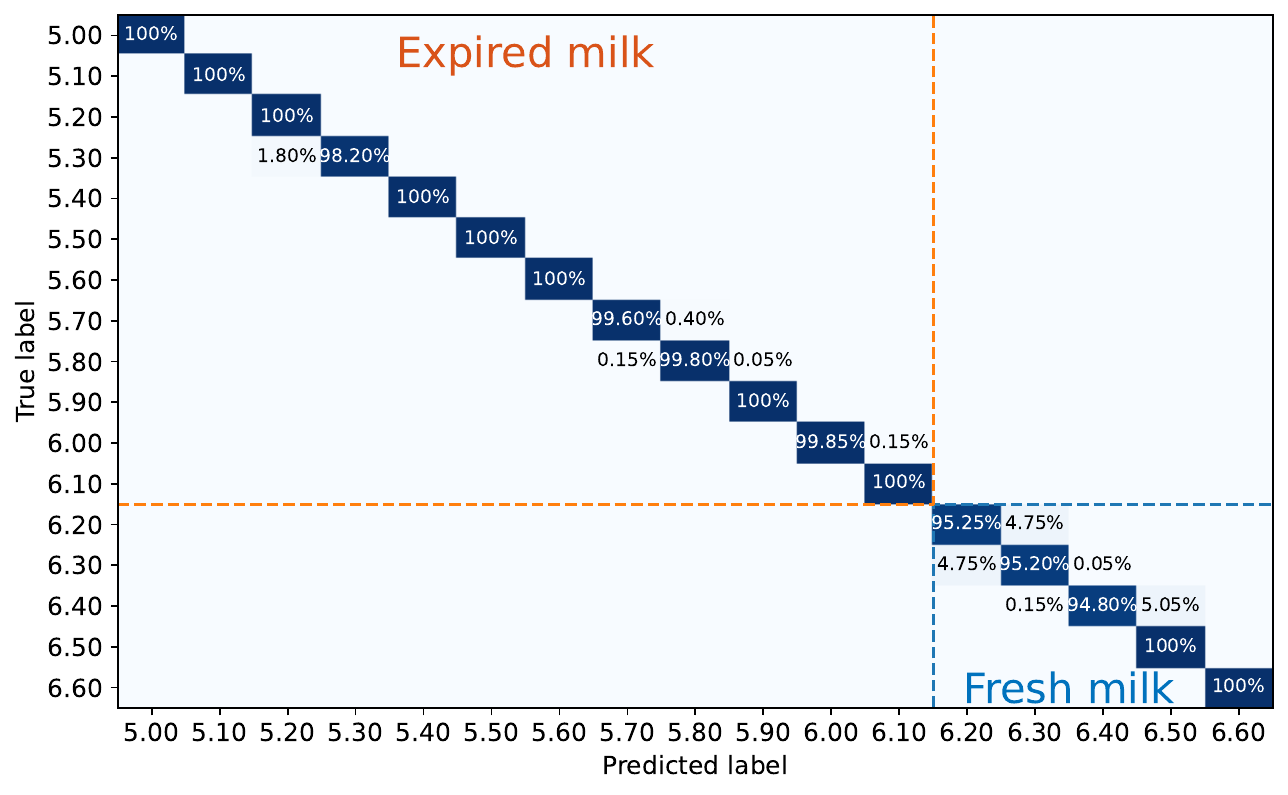}
}
\caption{Solution~(sugar and salt) concentration and milk pH level estimation accuracy.}
\end{figure*}

To detect milk spoilage, we divide the spoilage process into 16 stages, from fresh (pH = 6.6) to fully spoiled (pH =~5.0). A pH probe is used to monitor the ground truth pH level of the milk during the spoilage process.
The pH probe is put into the milk and the initial pH level of the fresh milk is 6.6.  
Specifically, we collect the sensing data at 17 pH levels, i.e., pH = 6.6, 6.5, 6.4, ..., 5.0. The pH level classification performance is presented in Figure~\ref{fig:ph_level}. When the pH level of the milk falls below 6.2, it typically indicates milk spoilage due to bacterial degradation. Our system can detect milk expiration at an accuracy of 100\%. 



\section{Conclusion}
In summary, this paper is the first attempt to utilize quantum wireless sensing to achieve unparalleled sensing accuracy for liquid recognition, taking advantage of the low noise and multi-frequency reception capability of a quantum receiver. Through light-weight learning, the proposed system can accurately differentiate highly similar liquids, measure fine-grained solution concentration and milk pH levels. Our experiments show that the proposed system can recognize seventeen liquids including very similar ones such as Pepsi and Coke at an accuracy higher than 99.9\%. The proposed system can detect a subtle 1\% change in sugar/salt concentration. Our system can sense milk pH levels at a granularity of 0.1, which is much finer than that required for expiration monitoring.


\bibliographystyle{IEEEtran} 
\bibliography{IEEEtran}                             

\begin{thebibliography}{10}
\providecommand{\url}[1]{#1}
\csname url@samestyle\endcsname
\providecommand{\newblock}{\relax}
\providecommand{\bibinfo}[2]{#2}
\providecommand{\BIBentrySTDinterwordspacing}{\spaceskip=0pt\relax}
\providecommand{\BIBentryALTinterwordstretchfactor}{4}
\providecommand{\BIBentryALTinterwordspacing}{\spaceskip=\fontdimen2\font plus
\BIBentryALTinterwordstretchfactor\fontdimen3\font minus \fontdimen4\font\relax}
\providecommand{\BIBforeignlanguage}[2]{{%
\expandafter\ifx\csname l@#1\endcsname\relax
\typeout{** WARNING: IEEEtran.bst: No hyphenation pattern has been}%
\typeout{** loaded for the language `#1'. Using the pattern for}%
\typeout{** the default language instead.}%
\else
\language=\csname l@#1\endcsname
\fi
#2}}
\providecommand{\BIBdecl}{\relax}
\BIBdecl

\bibitem{url4}
``The state of wireless.'' 2018, \url{https://api.ctia.org/wpcontent/uploads/2018/07/CTIA\_State-of-Wireless-2018\_0710.pdf}.

\bibitem{url3}
``Internet traffic from mobile devices.'' 2023, \url{https://explodingtopics.com/blog/mobile-internet-traffic}.

\bibitem{whole2013}
\BIBentryALTinterwordspacing
Q.~Pu, S.~Gupta, S.~Gollakota, and S.~Patel, ``Whole-home gesture recognition using wireless signals,'' in \emph{Proceedings of the 19th Annual International Conference on Mobile Computing $\&$ Networking}, ser. MobiCom '13.\hskip 1em plus 0.5em minus 0.4em\relax New York, NY, USA: Association for Computing Machinery, 2013, p. 27–38. [Online]. Available: \url{https://doi.org/10.1145/2500423.2500436}
\BIBentrySTDinterwordspacing

\bibitem{From2018}
F.~Zhang, D.~Zhang, J.~Xiong, H.~Wang, K.~Niu, B.~Jin, and Y.~Wang, ``From fresnel diffraction model to fine-grained human respiration sensing with commodity wi-fi devices,'' \emph{Proc. ACM Interact. Mob. Wearable Ubiquitous Technol.}, vol.~2, no.~1, Mar. 2018.

\bibitem{Fan2015}
\BIBentryALTinterwordspacing
H.~Fan, S.~Kumar, J.~Sedlacek, H.~Kübler, S.~Karimkashi, and J.~P. Shaffer, ``Atom based {RF} electric field sensing,'' \emph{Journal of Physics B: Atomic, Molecular and Optical Physics}, vol.~48, no.~20, p. 202001, sep 2015. [Online]. Available: \url{https://doi.org/10.1088/0953-4075/48/20/202001}
\BIBentrySTDinterwordspacing

\bibitem{gallagher1994rydberg}
\BIBentryALTinterwordspacing
T.~F. Gallagher, \emph{Rydberg {{Atoms}}}, ser. Cambridge {{Monographs}} on {{Atomic}}, {{Molecular}} and {{Chemical Physics}}.\hskip 1em plus 0.5em minus 0.4em\relax {Cambridge University Press}, 1994. [Online]. Available: \url{https://www.cambridge.org/core/books/rydberg-atoms/B610BDE54694936F496F59F326C1A81B}
\BIBentrySTDinterwordspacing

\bibitem{Wang:23}
\BIBentryALTinterwordspacing
Z.~Wang, M.~Jing, P.~Zhang, S.~Yuan, H.~Zhang, L.~Zhang, L.~Xiao, and S.~Jia, ``Noise analysis of the atomic superheterodyne receiver based on flat-top laser beams,'' \emph{Opt. Express}, vol.~31, no.~12, pp. 19\,909--19\,917, Jun 2023. [Online]. Available: \url{https://opg.optica.org/oe/abstract.cfm?URI=oe-31-12-19909}
\BIBentrySTDinterwordspacing

\bibitem{Jiao2017}
\BIBentryALTinterwordspacing
Y.~Jiao, L.~Hao, X.~Han, S.~Bai, G.~Raithel, J.~Zhao, and S.~Jia, ``Atom-based radio-frequency field calibration and polarization measurement using cesium $n{D}_{J}$ floquet states,'' \emph{Phys. Rev. Applied}, vol.~8, p. 014028, Jul 2017. [Online]. Available: \url{https://link.aps.org/doi/10.1103/PhysRevApplied.8.014028}
\BIBentrySTDinterwordspacing

\bibitem{Jing2020}
\BIBentryALTinterwordspacing
M.~Jing, Y.~Hu, J.~Ma, H.~Zhang, L.~Zhang, L.~Xiao, and S.~Jia, ``Atomic superheterodyne receiver based on microwave-dressed rydberg spectroscopy,'' \emph{Nature Physics}, vol.~16, pp. 911--915, 2020. [Online]. Available: \url{https://doi.org/10.1038/s41567-020-0918-5}
\BIBentrySTDinterwordspacing

\bibitem{Mass2012}
P.~Donato, F.~Cacciola, P.~Q. Tranchida, P.~Dugo, and L.~Mondello, ``Mass spectrometry detection in comprehensive liquid chromatography: basic concepts, instrumental aspects, applications and trends,'' \emph{Mass Spectrom Rev}, vol. 31(5), pp. 523--59, 2012.

\bibitem{LiquID2018}
A.~Dhekne, M.~Gowda, Y.~Zhao, H.~Hassanieh, and R.~R. Choudhury, ``Liquid: A wireless liquid identifier,'' in \emph{Proceedings of the 16th Annual International Conference on Mobile Systems, Applications, and Services}, ser. MobiSys '18.\hskip 1em plus 0.5em minus 0.4em\relax New York, NY, USA: Association for Computing Machinery, 2018, p. 442–454.

\bibitem{deng2019noninvasive}
J.~Deng, W.~Sun, L.~Guan, N.~Zhao, M.~B. Khan, A.~Ren, J.~Zhao, X.~Yang, and Q.~H. Abbasi, ``Noninvasive suspicious liquid detection using wireless signals,'' \emph{Sensors}, vol.~19, no.~19, p. 4086, 2019.

\bibitem{sedlacek2012microwave}
J.~A. Sedlacek, A.~Schwettmann, H.~K{\"u}bler, R.~L{\"o}w, T.~Pfau, and J.~P. Shaffer, ``Microwave electrometry with rydberg atoms in a vapour cell using bright atomic resonances,'' \emph{Nature Physics}, vol.~8, no.~11, pp. 819--824, 2012.

\bibitem{feynman2011feynman}
R.~P. Feynman, R.~B. Leighton, and M.~Sands, \emph{The Feynman lectures on physics, Vol. I: The new millennium edition: mainly mechanics, radiation, and heat}.\hskip 1em plus 0.5em minus 0.4em\relax Basic books, 2011, vol.~1.

\bibitem{Mohapatra2007}
\BIBentryALTinterwordspacing
A.~K. Mohapatra, T.~R. Jackson, and C.~S. Adams, ``Coherent optical detection of highly excited rydberg states using electromagnetically induced transparency,'' \emph{Phys. Rev. Lett.}, vol.~98, p. 113003, Mar 2007. [Online]. Available: \url{https://link.aps.org/doi/10.1103/PhysRevLett.98.113003}
\BIBentrySTDinterwordspacing

\end{thebibliography}


\begin{IEEEbiographynophoto}{Yuechun Jiao}
(ycjiao@sxu.edu.cn) received his Ph.D. degree in Optics from Shanxi University, Taiyuan, China, in 2017. He is currently an associate professor at State Key Laboratory of Quantum Optics and Quantum Optics Devices, Institute of Laser Spectroscopy, Shanxi University. His research interests include Rydberg quantum optics and Microwave sensors based on Rydberg atoms.
\vspace{-3em}
\end{IEEEbiographynophoto}
\begin{IEEEbiographynophoto}{Jinlian Hu}
is currently working toward a Ph.D. degree in Atomic and Molecular Physics from  Institute of Laser Spectroscopy, Shanxi University. Her research interests include microwave sensors based on Rydberg atoms.
\vspace{-3em}
\end{IEEEbiographynophoto}
\begin{IEEEbiographynophoto}{Zitong Lan}
is currently working toward a Ph.D. degree at ESE department of University of Pennsylvania. His research interests include wireless sensing, mobile and ubiquitous computing.
\vspace{-3em}
\end{IEEEbiographynophoto}
\begin{IEEEbiographynophoto}{Fusang Zhang}
(fusang@iscas.ac.cn) received the M.S. and Ph.D. degrees
in computer science from the Institute of Software,
Chinese Academy of Sciences, Beijing, China, in
2013 and 2017, respectively. He is currently an
Associate Professor with the Institute of Software,
Chinese Academy of Sciences. His current research
interests include mobile and pervasive computing, wireless  sensing and quantum sensing.
\vspace{-3em}
\end{IEEEbiographynophoto}
\begin{IEEEbiographynophoto}{Jie Xiong} received his PhD degree in Computer Science from University College London in 2015. He is currently a Principal Researcher at Microsoft Research Asia an also an Associate Professor at the College of Information and Computer
Sciences, University of Massachusetts Amherst. His recent research interests include wireless sensing and mobile computing.
\vspace{-3em}
\end{IEEEbiographynophoto}
\begin{IEEEbiographynophoto}{Jingxu Bai}
is currently working toward a Ph.D. degree in Atomic and Molecular Physics from  Institute of Laser Spectroscopy, Shanxi University. Her research interests include microwave sensors based on Rydberg atoms.
\vspace{-3em}
\end{IEEEbiographynophoto}
\begin{IEEEbiographynophoto}{Zhaoxin Chang}
is currently working toward a Ph.D. degree in Telecom SudParis, Institut Polytechnique de Paris. His research interests include wireless sensing, mobile and ubiquitous computing.
\vspace{-3em}
\end{IEEEbiographynophoto}
\begin{IEEEbiographynophoto}{Yuqi Su}
is currently pursuing the master degree in Institute of Software,
Chinese Academy of Sciences, Beijing. Her research interests include
mobile and pervasive computing, wireless sensing, and quantum sensing.
\vspace{-3em}
\end{IEEEbiographynophoto}

\begin{IEEEbiographynophoto}{Beihong Jin} received the BS degree in computer
science from Tsinghua University, in 1989, and the MS and PhD degrees in computer science from the Institute of Software, Chinese Academy of Sciences, in 1992 and 1999, respectively. Currently, she is a full professor in the Institute of Software, Chinese Academy of Sciences. Her research interests include mobile and pervasive computing, recommended system, and wireless sensing.
\vspace{-3em}
\end{IEEEbiographynophoto}

\begin{IEEEbiographynophoto}{Daqing Zhang} (Fellow, IEEE) received the Ph.D.
degree from the University of Rome “La Sapienza,”
Italy, in 1996. He is currently a Chair Professor
with the School
of Computer Science, Peking University, China; and
Telecom SudParis, IP Paris, France. His research interests include
context-aware computing, wireless sensing, and quantum sensing.
\vspace{-3em}
\end{IEEEbiographynophoto}

\begin{IEEEbiographynophoto}{Jianming Zhao}
(zhaojm@sxu.edu.cn) received her Ph.D. degree in Optics from Shanxi University, Taiyuan, China, in 2003. She is currently a professor at State Key Laboratory of Quantum Optics and Quantum Optics Devices, Institute of Laser Spectroscopy, Shanxi University. Her research interests include ultracold atomic and molecular Physics and Microwave sensors based on
Rydberg atoms.
\vspace{-3em}
\end{IEEEbiographynophoto}

\begin{IEEEbiographynophoto}{Suotang Jia}
received his Ph.D. degree in Optics from East China Normal University, Shanghai, China, in 1994. He is currently a professor at Institute of Laser Spectroscopy, Shanxi University.
\vspace{-3em}
\end{IEEEbiographynophoto}

\vfill

\end{document}